Letter

# Re-investigation of the interplay of fission modes and non-equilibrium fission processes in heavy actinide nuclei $^{249}$Bk and $^{257}$Md

R. Dubey [a,*], G. Kaur [b], Abhishek Yadav [c], N. Saneesh [d], Akhil Jhingan [d], P. Sugathan [d], Tathagata Banerjee [e], K.S. Golda [c], Hardev Singh [f], Meenu Thakur [g], Ruchi Mahajan [h]

[a] Institute of Physics, University of Szczecin, 70-451 Szczecin, Poland
[b] CEA Irfu, Université Paris-Saclay, F-91191 Gif-sur-Yvette, France
[c] Amity Institute of Nuclear Science and Technology, Amity University, Noida-201313, India
[d] Inter-University Accelerator Centre, Aruna Asaf Ali Marg, New Delhi - 110067, India
[e] Istituto Nazionale di Fisica Nucleare, Sezione di Napoli, 80126 Napoli, Italy
[f] Department of Physics, Kurukshetra University, Kurukshetra, Haryana-136119, India
[g] Department of Physics and Astrophysics, Central University of Haryana, Jant-Pali, Mahendergarh, Haryana-123031, India
[h] Department of Physics and Astronomy, Louisiana State University, Baton Rouge, LA 70803, USA

ARTICLE INFO

ABSTRACT

Measurements of mass and angular distributions of fission fragments from actinide nuclei $^{249}$Bk and $^{257}$Md, produced in fusion reactions $^{11}$B and $^{19}$F + $^{238}$U, are presented. Experimentally observed mass ratio distributions indicate "multi-chance fission" through the interplay of fission modes in the fission process, and they agree well with predictions from calculations using the GEF ("GEneral description of Fission observables") model code. Furthermore, to test the signatures of events from non-compound nuclear processes in the fission of $^{249}$Bk and $^{257}$Md nuclei, Monte Carlo statistical decay model calculations using GEMINI$^{++}$ were performed for the measured mass distribution at all energies. For comparison purposes, the fission fragment mass distributions of neighboring heavy actinide nuclei, previously measured in the fission of $^{250}$Cf and $^{254}$Fm nuclei produced by $^{12}$C and $^{16}$O projectiles on a $^{238}$U target, are also presented. The measured angular anisotropy data for the $^{19}$F + $^{238}$U reaction differ from the results of the Transition State Model (TSM) at energies below the fusion barrier. As a result of the present study, we suggest considering the interplay between K relaxation time, dynamic dissipation, and their influence on shell correction to understand the evolution of fission dynamics in heavy-ion-induced actinide nuclei.

## 1. Introduction

Heavy elements are synthesized in the laboratory through heavy ion fusion reactions. It is expected that two nuclei overcome the Coulomb repulsion and fuse to form a statistically equilibrated compound nucleus (CN) that survives fission, leading to evaporation residues (ER). However, for heavier colliding partners, the strong Coulomb repulsion between the interacting nuclei makes it more difficult to form a fully equilibrated heavy actinide or super heavy CN, and the fusion-fission dynamics become more complex due to the interplay of many entrance channel parameters. Presently, studies investigating the fission of heavy actinide nuclei have revealed the complex nature of fission dynamics as evidenced by anomalous increase in fission fragment (FF) mass variance and angular anisotropies [1,2].

The role of entrance channel parameters is more evident in fusion-fission reactions using heavy projectiles (i.e., mass number $A_p$ greater than 22) bombarding on actinide targets, as they show strong dominance of orientation-dependent quasi-fission (QF) in the sub-barrier energy regime [3,4,6,5,7]. In fusion-fission reactions involving light mass projectiles such as $^{12}$C, $^{16}$O, and $^{19}$F bombarding actinide targets, the anomalous increase in angular anisotropies were attributed to the onset of pre-equilibrium fission (PEF) [8,9,11–13] or slow QF [10,14]. However, for such reactions, ambiguity still exist for describing the anomalous increase in FF mass variance and angular anisotropies observed at sub-barrier energies. For instance, in reactions such as $^{16}$O + $^{238}$U and $^{12}$C + $^{236}$U, though anomalous angular anisotropies were observed [12], no ER suppression or any mass angle correlation was reported [15]. ER measurement concluded complete fusion observed even at deep sub barrier energies in $^{16}$O + $^{238}$U [15]. In the decay of compound nucleus $^{254}$Fm formed through two different reaction channels: $^{11}$B + $^{243}$Am and $^{16}$O + $^{238}$U, despite observing large angular anisotropies in both reactions, mass distribution ruled out QF in former reaction [16]. Although entrance channel mass asymmetry and target deformations have been attributed to this anomalous behavior, it is obvious that in the induced fission of actinide targets with low charge product $Z_P Z_T$ ($Z_P$ and $Z_T$ are the atomic charge number of projectile and target nuclei, respectively), identification of QF is ambiguous.

* Corresponding author.
*E-mail address:* rakesh.dubey@usz.edu.pl (R. Dubey).

Recently, the role of multi-chance fission (MCF) and different fission modes have been examined to investigate the mass distribution and total kinetic energy distribution [17–19]. Through MCF, fission preceded by neutron emission reduces the initial excitation energy ($E^*$) which can lead to different fission modes, including asymmetric (AS) and symmetric superlong (SL), supershort (SS) fission modes [20–22]. Generally, these fission modes are discussed in terms of deviations from the Gaussian nature of mass and total kinetic energy (TKE) distributions [23]. In a recent work [24], based on TKE distributions observed in fusion-fission reaction $^{19}$F + $^{238}$U $\longrightarrow$ $^{257}$Md, bimodal fission has been reported in $^{257}$Md with supershort mode persistent even up to $E^* \approx 48$ MeV. However, the observed broadening of FF mass distribution was treated as features of slow QF with shell structures of fragments influencing the mass equilibration process [25]. As shell effects in measured mass distributions are described well invoking MCF [26] and different fission modes are discussed as properties of fully equilibrated CN, investigating the role of fission modes and non-equilibrium fission in similar systems through FF angular anisotropies and mass distributions for a range of excitation energies would be of great interest.

We have measured the FF angular distribution for reaction $^{19}$F + $^{238}$U and mass distributions for reactions $^{11}$B, $^{19}$F + $^{238}$U around fusion barrier energies. Preliminary results were presented in a conference proceeding [27]. In this letter, we present the analysis of observed near symmetric mass distributions in both systems $^{11}$B, $^{19}$F + $^{238}$U but rather anomalous angular anisotropy in reaction $^{19}$F + $^{238}$U. For the reaction $^{11}$B + $^{238}$U, the fission fragment angular distributions were considered based on the work of Liu et al. and Karnik et al. [11,28] The angular anisotropies were analyzed within the framework of transition-state model (TSM) [29] and compared with measurements of Ref. [11,28]. The fission modes and non-equilibrium fission processes are explored in measured mass distributions by comparing them with predictions from the GEF code [18] and the statistical model code GEMINI$^{++}$ [30]. By analyzing both the mass distribution and angular distribution for $^{257}$Md and $^{249}$Bk nuclei, we attempt to make a careful assessment of the contributions of non-compound events and any possible interplay between K relaxation time and different fission modes.

## 2. Experimental details

The experiments were performed using $^{11}$B and $^{19}$F ion beams from the 15UD Pelletron accelerator at Inter-University Accelerator Centre (IUAC), New Delhi. Enriched $^{238}$U target of thickness ∼ 110 μg/cm$^2$, sandwiched between Carbon backings (∼ 20 μg/cm$^2$) was used. For mass distribution measurements, pulsed beams with ∼ 1.4 nanoseconds (ns) bunch width separated by 250 ns were employed. Beam energies were varied from 92 MeV to 120 MeV for $^{19}$F and 52 to 66 MeV for $^{11}$B respectively. The experimental setup for mass distribution measurement consisted two large area ($20cm \times 10cm$) Multi-Wire Proportional Counters (MWPCs) mounted on two arms inside the 1.5 m diameter scattering chamber [32]. The detectors were kept asymmetrically, one at forward angle ($\theta_{f1} = 40^o$ at 35 cms) and the other at backward angle ($\theta_{f2} = 130^o$ at 27 cms) with respect to beam direction. Fission products were detected in coincidence and the time of flight measured with respect to the R.F signal of the beam pulse. In a separate experiment, FF angular distributions were measured for the reaction $^{19}$F + $^{238}$U using an array of ΔE-E gas-silicon telescope detectors [33]. Angular distributions were measured in the laboratory angles ranging from 85° to 175° with respect to the beam direction. Two silicon monitor detectors were mounted at forward angles to enable the determination of absolute fission cross sections. The reaction parameters are given in Table 1.

**Table 1**
Main parameters include the compound nucleus (CN) populated by the reaction, the capture barrier ($V_B$), product of charges of colliding partners($Z_P Z_T$), beam energy in the lab frame ($E_{lab}$), and corresponding excitation energies ($E^*$) of CN for the studied reactions $^{11}$B, $^{19}$F + $^{238}$U.

| Reactions | CN | $V_B$ (MeV) | $Z_P Z_T$ | $E_{lab}$ (MeV) | $E^*$ (MeV) |
|---|---|---|---|---|---|
| $^{11}$B + $^{238}$U | $^{249}$Bk | 53.67 | 460 | 50 | 33.91 |
| | | | | 52 | 35.83 |
| | | | | 54 | 37.78 |
| | | | | 58 | 41.56 |
| | | | | 62 | 45.38 |
| | | | | 64 | 47.38 |
| | | | | 66 | 49.21 |
| $^{19}$F + $^{238}$U | $^{257}$Md | 93.67 | 828 | 92 | 42.00 |
| | | | | 94 | 43.87 |
| | | | | 98 | 47.58 |
| | | | | 101 | 50.55 |
| | | | | 108 | 56.84 |
| | | | | 112 | 60.54 |
| | | | | 115 | 63.32 |
| | | | | 120 | 67.95 |

## 3. Results and discussion

### 3.1. Mass distribution

Analysis of mass distribution was performed using event by event reconstruction of the two-body kinematics in the center-of-mass system [10]. For each coincident event data, the MWPC position (X,Y) was transformed to extract the scattering angle $\theta$ and the azimuthal angle $\phi$ for both fragments. The velocity vector components, projected onto plane perpendicular to the beam ($V_\perp$) and parallel to the beam axis ($V_\parallel$) were extracted. Full momentum transferred fission events were separated by applying software gates on events for which distribution of $V_\perp$ and $V_\parallel - V_{c.m.}$ ($V_{c.m.}$ is the calculated center-of mass velocity) peaks at zero. The mass-angle and the mass ratio $M_R = \frac{m_1}{m_1+m_2}$ (where $m_1$ and $m_2$ are masses of two primary fragments) distributions were generated at different beam energies for the two reactions $^{11}$B, $^{19}$F + $^{238}$U. Though this method has been commonly used to analyze experimental mass distribution, significant uncertainty in the TKE distribution of fission fragments might arise due to traversal through a medium before the final detection of fission fragments. As a consequence, it presents challenges in extracting precise TKE distribution and constrains clear interpretation of reaction mechanisms, such as multi-modal fission. Hence in present work, we will try to understand fission mechanism with help of mass-ratio and angular anisotropy distributions of FF.

Fig. 1(left panel) displays the measured FF mass-ratio distributions for the two reactions, $^{11}$B + $^{238}$U (upper panel) and $^{19}$F + $^{238}$U (lower panel), at three different excitation energies, as indicated for each measurement. First, we attempted to estimate the contributions of non-equilibrium events in the mass distributions for $^{249}$Bk and $^{257}$Md nuclei by comparing the experimental mass distributions with those predicted by the GEF model, shown as the grey-shaded region representing symmetric components. We then subtracted the calculated distributions from the experimental ones, presented as the magenta-shaded region representing asymmetric components. The difference between the extracted distributions and GEF predictions could be attributed to the presence of slow quasi fission events [10,14], or fission after the seventh or very higher chances of fission [1]. However, the measured distributions were found to be nearly symmetric and could be well represented by a near-Gaussian shape around $M_R = 0.5$ in all cases. These observations inidicate that the FF mass distribution corresponds mostly to complete fusion-fission events in both reactions.

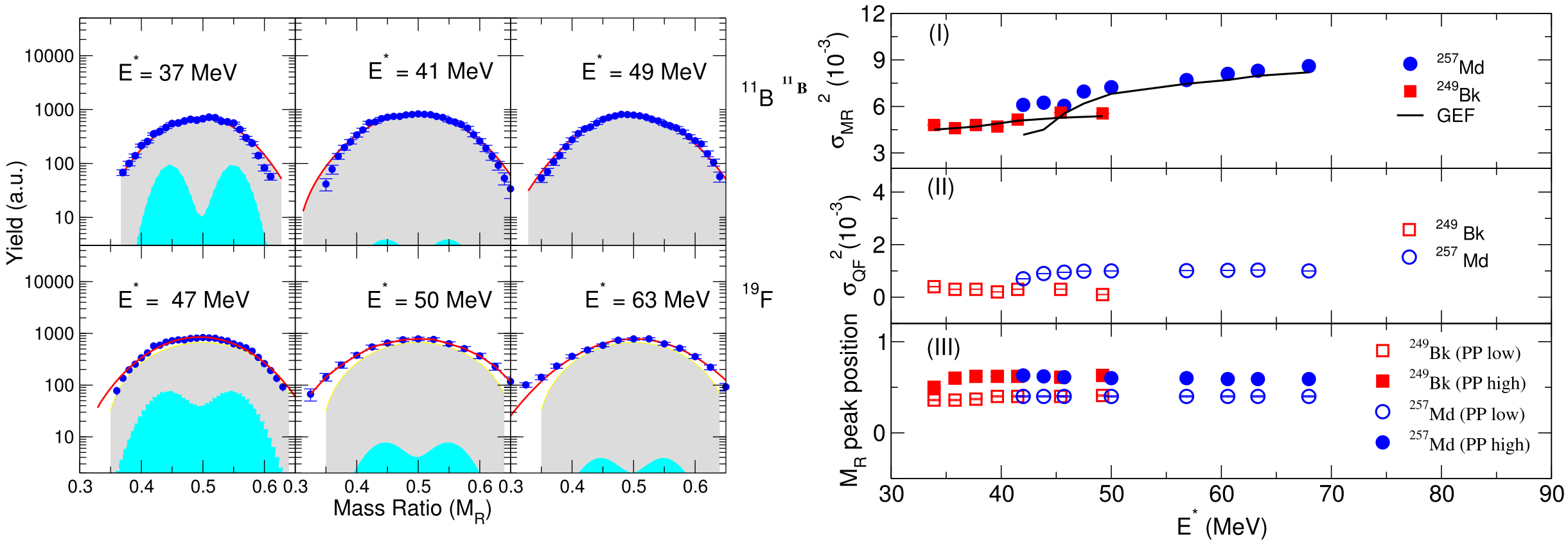

**Fig. 1.** Left panel: Figures show the mass-ratio distributions blue filled circles of fission fragments from $^{11}B$, $^{19}F + ^{238}U$ reactions forming $^{249}Bk$, $^{257}Md$ nuclei. The results were compared with GEF calculations, shown in grey shaded regions. The magenta shaded regions represent the difference between GEF model calculations and measurements and red solid curve corresponds to Gaussian fit to GEF total (broadened). Right panel: Figures show the experimental and GEF predicted mass-ratio variance [I], QF mass variance estimated from mass asymmetric components (magenta shaded region) [II], and mass-ratio peak position (PP) for lighter (PP low) and heavier fragments (PP high) [III] as a function of excitation energies for the fission of $^{257}Md$ and $^{249}Bk$ nuclei.

Furthermore, if we examine the extracted $M_R$ widths compared with those obtained from GEF results, as shown in Fig. 1(right panel), it is evident that the mass ratio widths for $^{249}Bk$ are well described by the GEF predictions, though slight deviations are observed for $^{257}Md$ at lower excitation energies. In the procedure of extracting the $M_R$ widths, the $M_R$ peak positions corresponding to the light and heavy fission fragments are illustrated in Fig. 1(b)(III). For $^{249}Bk$, across all beam energies, the centroids of the lighter and heavier fragment peaks consistently occur at mass ratios of approximately 0.4 and 0.6, however, for $^{257}Md$, slight deviations are observed. This suggests that the lighter fragments might correspond to the closed-shell nucleus $^{96}Zr$, while the heavier fragments could be related to tin cluster formations. However, the exact identification of these fragments is challenging due to potential shell corrections influenced by the presence of multi-chance fission. To provide further insight into the multi-chance fission process in these nuclei, mass-ratio distributions were calculated for each fission chance, with dashed curves representing different fission chances (e.g., first-chance in magenta, third-chance in orange, fifth-chance in black), as shown in Fig. 2(a)). The reduction in excitation energy due to neutron emission was calculated using neutron binding energies and an average neutron energy of approximately 2.2 MeV, as determined by the PACE4 code. The overall FF mass ratio distributions, up to the fifth fission chance, is depicted by the thin black curve, which closely aligns with the experimentally observed peak positions, though with narrower peaks.

A key observation from Fig. 2(a) is that the mass-asymmetric fission seen in $^{257}Md$ at high excitation energies primarily arises from lower-energy higher-chance of fissions (e.g., $^{250}Md$). In contrast, fissions occurring during the first to fourth chances at higher excitation energies tend to result in symmetric mass splits. For highly excited $^{257}Md$, symmetric fission is predominant, while later fission chances, especially after the fifth chance in $^{252}Md$, display a trend toward asymmetric fission. Similarly, in $^{249}Bk$, the fourth fission chance in $^{244}Bk$ shows a dominant asymmetric fission pattern.

To determine the contribution from different fission modes, the measured distributions were fitted using Gaussian functions representing one symmetric component (grey-filled area) and several asymmetric components (dashed curve). We attempted to investigate the probability of the presence of other possible modes, such as the super-short mode, by subtracting the symmetric and asymmetric components from the GEF total broadened mass-ratio distribution. As shown in Fig. 2 (I, II) for both nuclei, the contribution from these other modes is negligibly small at excitation energies below and above 30 MeV. In contrast, the asymmetric mode, originating from a single fissioning nucleus, could play a significant role at $E^* < 30$ MeV. In the case of $^{257}Md$, the strength of asymmetric modes begins to appear even beyond 60 MeV excitation energies. However, it is challenging to distinguish between AS and SS modes using this approach, as the analysis is primarily based on GEF model predictions. In Fig. 2(c), the total contribution of the asymmetric modes for the $^{257}Md$, $^{249}Bk$, and $^{250}Cf$ actinide nuclei as a function of excitation energy is presented and compared with other works in the literature.

Even considering the comparatively small percentage of the SS mode as predicted by GEF calculations, where the most likely FF fragment nuclei ($Z \sim 56$, $A \sim 142$) or two doubly magic nuclei ($^{132}Sn$) do not significantly alter the total asymmetric components. In the case of spontaneous fission from $^{259,260}Md$, $^{258}Fm$, and $^{258}No$, we can clearly distinguish the SS and AS modes due to the strong shell effect at zero excitation energy [1,26]. However, the fission of these nuclei produced by heavy ion beams is quite different, as these nuclei are produced via a complex fusion process that evolves until fission occurs, even at the lowest possible excitation energies. Thus, the direct comparison between AS and SS fission modes in spontaneous fission and heavy-ion-induced fusion fission for heavy actinide nuclei is not justifiable.

### 3.2. Angular distribution

The FF angular distributions for the reaction $^{19}F + ^{238}U$ were transformed into the center-of-mass frame assuming the Viola systematics [35] for symmetric fission. The measured angular distributions, $W(\theta)$ were fitted using Legendre polynomials to extrapolate the data to $0^o$ or $180^o$ and deduce the FF angular anisotropy given by $A \equiv W(0^o or 180^o)/W(90^o)$. The anisotropies were calculated using the approximate relation;

$$A \approx 1 + \frac{\langle J^2 \rangle}{4K_0^2} \tag{1}$$

where $\langle J^2 \rangle$ is the mean squared angular momentum of the fissioning nuclei, and $K_0^2$ the variance of the K (projection of $J$ onto nuclear symmetry axis) distribution. The variance $K_0^2$ was calculated as $\frac{\Im_{eff} T}{\hbar^2}$, where $\Im_{eff}$ is effective moment of inertia and $T$ the temperature of the fissioning nucleus at the saddle point deformation. $\Im_{eff}$, was obtained from the rotating finite range model (RFRM) calculations [36].

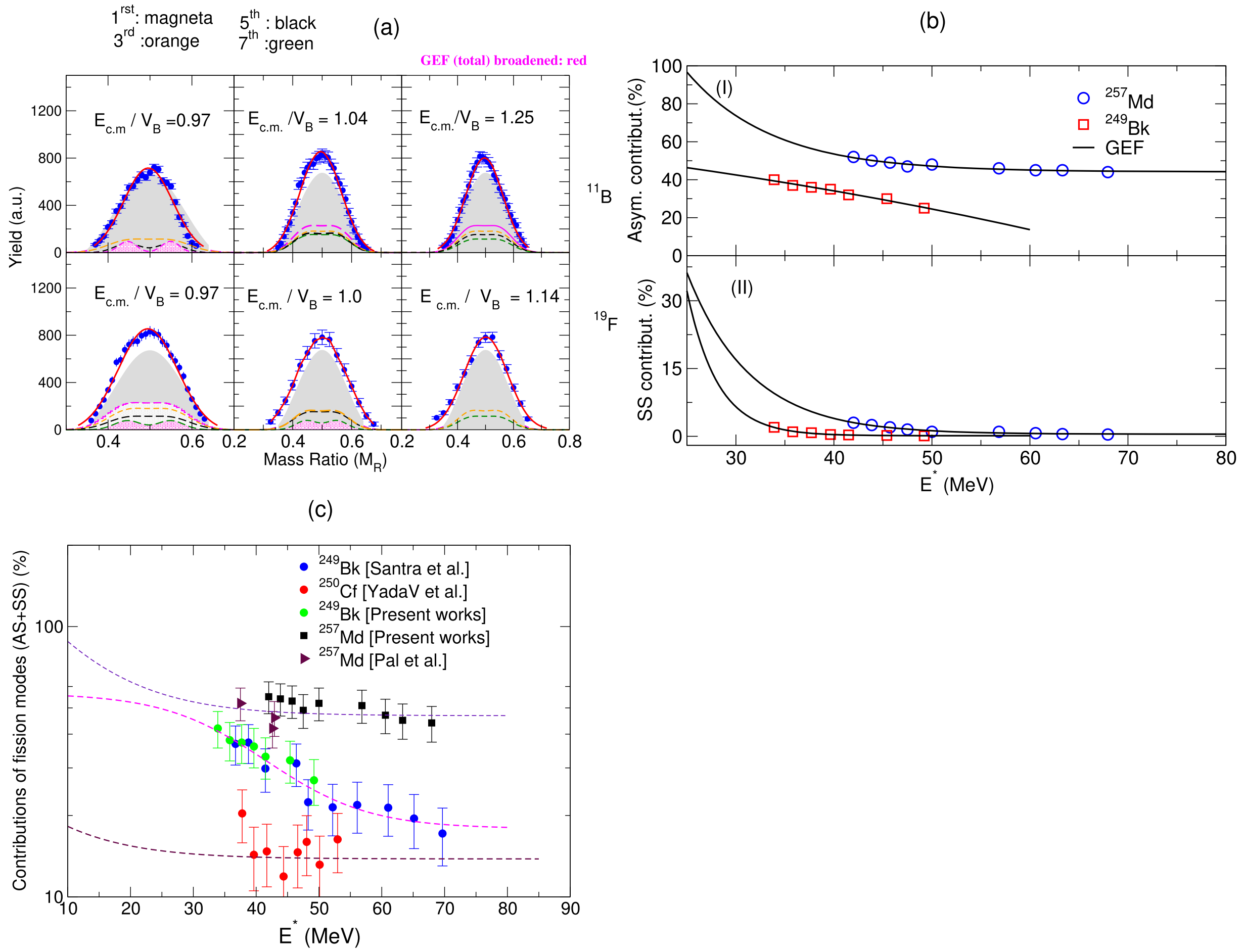

**Fig. 2.** (a) The figure shows the mass-ratio distributions of fission fragments from $^{11}B$ and $^{19}F + {}^{238}U$ reactions for three excitation energies of $^{249}Bk$ and $^{257}Md$ compound nuclei, compared with GEF calculations incorporating contributions from chance fissions (illustrated by broken lines of different colors). The columns (left to right) represent $\frac{E_{c.m.}}{V_B}$ corresponding to excitation energies below, at, and above the respective fusion barriers. The red lines depict GEF calculations broadened by experimental mass resolution, characterized by $\sigma = 7.5\,u$. (b) [I] The excitation energy dependence of asymmetric fission contributions (symbols), extracted from measured experimental mass distributions for $^{249}Bk$ and $^{257}Md$ actinide nuclei, is compared with GEF predictions (black solid curve). [II] Contributions of the other modes (in percentage) are presented as estimated after subtracting the symmetric and asymmetric components from the GEF total broadened fitted mass-ratio distribution. (c) Comparison of re-normalized contributions of different fission modes (SS + AS, symbols) with GEF calculations (dashed lines) for $^{249}Bk$, $^{250}Cf$, and $^{257}Md$. Some of the data points for the $^{249}Bk$, $^{257}Md$ and $^{250}Cf$ nuclei are taken from studies by Santra et al., Yadav et al., and Pal et al. [24,26,34].

The temperature at the saddle point was calculated after correction for pre-scission neutron emissions [37] and using the level density parameter ‘$a$’ = $A_{CN}/8.5$ MeV. The angular momentum distributions of the fissioning nuclei were calculated using the coupled channel code CCFULL [38] using established parameters that reproduced the fission cross sections.

The angle integrated fission cross sections as a function of $E_{c.m}$ for $^{19}F + {}^{238}U$ are shown in Fig. 3(a). Since CN $^{257}Md$ is highly fissile, we assume the fission cross section to be the same as total fusion cross section $\sigma_{fus}$. The fusion excitation functions generated using CCFULL are also shown in Fig. 3 (a). The measured cross sections are reasonably reproduced (solid curve) when prolate deformations of $^{238}U$ ($\beta_2 = 0.275$, $\beta_4 = 0.05$) and coupling to the first excitation at 0.73 MeV and $\beta_3 = 0.086$ for $^{238}U$ are included in the coupled-channel calculations as given in Ref. [15]. The dotted curve represents the calculated fusion cross section when deformation and coupling to collective states were not included in the coupling scheme. Without couplings the cross section drops below the Coulomb barrier showing the importance of nuclear properties on capture cross sections. Fig. 3(b) shows the experimental anisotropies deduced from the fit to angular distribution data for $^{19}F + {}^{238}U$ system in the present work. Also shown in the figure are data from Ref. [28] at few energies above barrier. The anisotropies calculated based on the TSM are shown by solid curve in Fig. 3(b). It is observed that measured angular anisotropies are large as compared to the TSM predictions and it rises rapidly with decrease in beam energy through fusion barrier region, in a manner similar to that observed for the reaction $^{16}O + {}^{238}U$ reported in [9,10]. It may be noted that, since angular distribution measurements are performed in singles, the measured data may contain contribution from transfer induced components. The present result is consistent with other measurements on actinide targets confirming the anomalously large fission anisotropies at sub barrier energies [8,10–13].

To gain further insight into $K$-equilibration in the fission process of these nuclei, the ratio $\langle J^2\rangle_{expt}/\langle J^2\rangle_{theory}$ as a function of $E_{c.m.}/V_B$ for systems with $\alpha \approx \alpha_{BG}$ is shown in Fig. 4(a). The theoretical values of $\langle J^2\rangle_{theory}$ were derived from a coupled-channels model (CCFULL) fit to experimental fusion-fission cross sections, while the experimental $\langle J^2\rangle_{expt}$ values were obtained from anisotropy measurements [11,28] based on $K_0^2$ values. As illustrated in Fig. 4(a), the experimental $\langle J^2\rangle_{expt}$ values align well with the coupled-channels model for $^{249}Bk$, but for $^{257}Md$, the experimental $\langle J^2\rangle_{expt}$ values are significantly higher than the theoretical predictions.

From this comparison, it appears that at lower angular momentum for systems where $\alpha < \alpha_{BG}$, the relaxation time of $K$-degrees of freedom may exceed the fission lifetime. Although several studies have addressed these observations, a notable approach by Liu et al. considered the relaxation process of $K$-degrees of freedom. The variance of the $K$-distribution, $\sigma_K^2$, is expressed as [11]:

$$\sigma_K^2 = K_0^2 \left[1 - \exp\left(-\frac{t}{\tau_K}\right)\right], \tag{2}$$

where $\tau_K$ is the relaxation time and $K_0^2$ is the equilibrium value of $\sigma_K^2$, as given by Eq. (2). Døssing and Randrup investigated the dynamic evolution of angular momentum during nuclear reactions and derived equations governing the evolution of the $K$-distribution [39]. They presented an expression for the relaxation time of $K$-degrees of freedom, dependent on rotational frequency $\omega_R$. Using certain approximations, the following equation for pre-equilibrium fission was obtained:

$$\sigma_K^2(J) = K_0^2 \left[1 - \exp\left(-GJ^2\right)\right], \tag{3}$$

with $G = \frac{2.238 g_I^2}{G_1^2 G_{\text{eff}}}$.

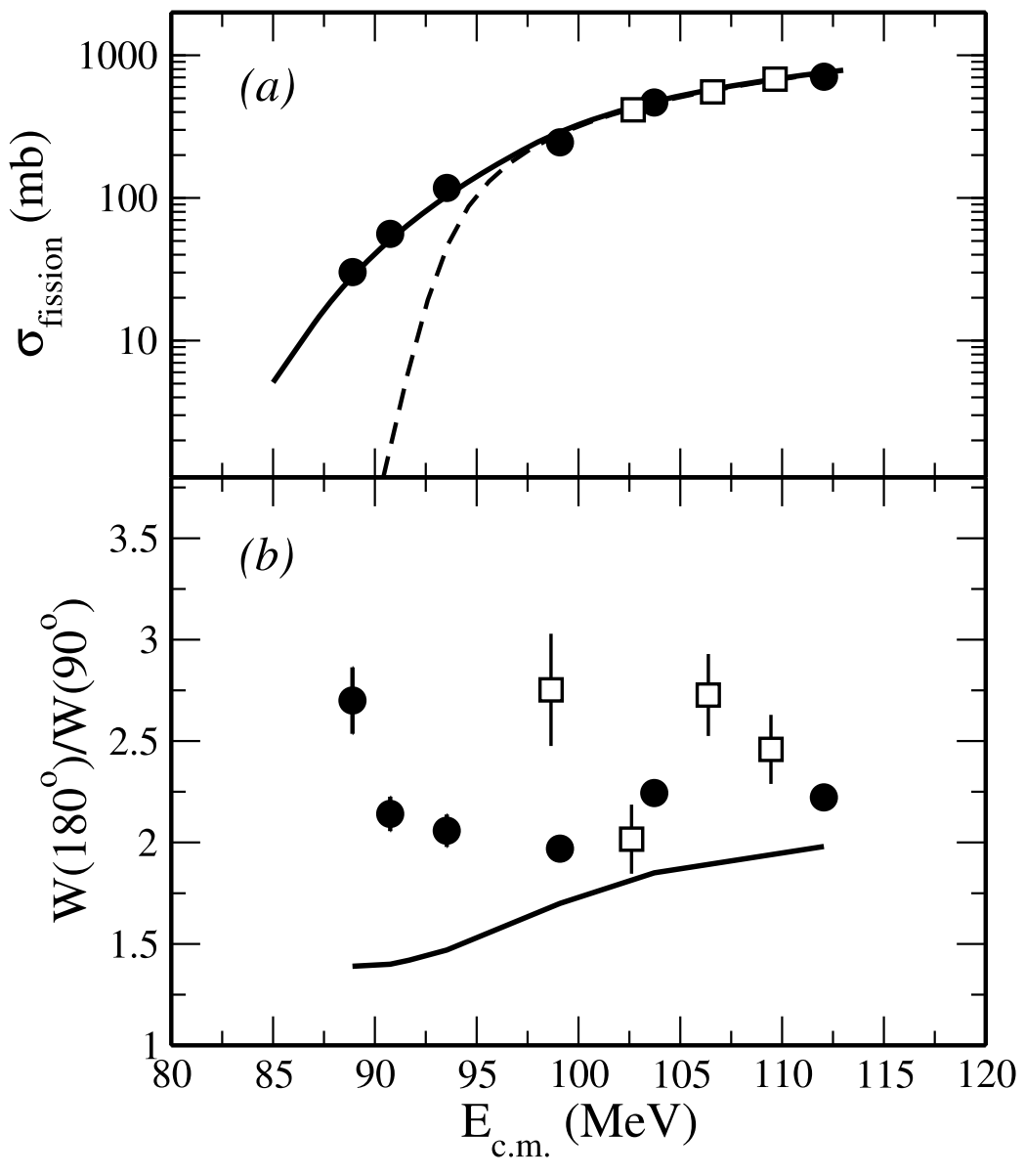

**Fig. 3.** Measured fission cross-section as a function of $E_{c.m}$ for $^{19}$F + $^{238}$U along with model calculated excitation function (solid curve). The dashed line shows calculation without any couplings (see text). (b) Fission anisotropies (filled circles) as function of $E_{c.m}$. The curve is the TSM calculation (see text). The open squares are data from Karnik et al. works [Ref [28]].

The average value of $\frac{\sigma_K^2}{K_0^2}$ is defined as:

$$\left\langle \frac{\sigma_K^2}{K_0^2} \right\rangle = \frac{\sum_{J=0}^{\infty} \sigma_F(J) \left[\frac{\sigma_K^2(J)}{K_0^2}\right]}{\sum_{J=0}^{\infty} \sigma_F(J)}. \tag{4}$$

Using these equations, the ratio $\frac{\sigma_K^2(J)}{K_0^2}$ was calculated and plotted as a function of mean-square angular momentum $\langle J^2 \rangle$.

As illustrated in Fig. 4(b), the $K$-degrees of freedom do not achieve complete equilibration at lower angular momentum in fusion process of $^{19}$F +$^{238}$ U reaction. The alignment between experimental results and theoretical predictions suggests the exponential relationship of $\sigma_K^2$ with $J^2$. When $J^2$ increases, $\tau_K$ diminishes sharply, enabling the $K$-degrees of freedom to reach equilibrium by the saddle point stage. Integrating the relaxation of $K$-degrees of freedom within pre-equilibrium fission also leads to the conclusion, consistent with Liu et al. [11], that coupled-channels calculations can effectively model both fusion excitation functions and angular momentum distributions, regardless of whether the fission process is non-equilibrium or fully equilibrated. Moreover, under conditions of low angular momentum, they proposed that the relaxation time of the $K$-degrees of freedom surpasses the fission timescale. Therefore, studies of mass and angular distributions in both nuclei imply a possible correlation between $K$ relaxation times and the fission pathways in the formation and disintegration of heavy actinide nuclei around the mass 250 region.

Furthermore, we extended our mass distribution studies for fission induced by projectiles $^{11}$B, $^{12}$C, $^{16}$O, $^{19}$F, and $^{30}$Si bombarding a $^{238}$U target using the Monte Carlo statistical decay model code GEMINI$^{++}$ [30], with support from existing studies on neighboring nuclei. Compared to GEF, GEMINI$^{++}$ is one of the most widely used codes for simulating the decay of the compound nucleus, following the Hauser-Feshbach formalism for particle evaporation and the Bohr-Wheeler formalism for the fission process [31]. It provides better insights into the interplay between fusion dynamics and fission outcomes, making it a superior choice for studying the influence of excitation energy and angular momentum on the final fission fragment mass distribution. The model assumes equilibration of all degrees of freedom associated with fission.

In Fig. 5, we present the fission fragment mass-ratio distributions from the current measurements, compared with data from other reactions using $^{12}$C, $^{16}$O, and $^{30}$Si projectiles on a $^{238}$U target along with the corresponding entrance channel parameters presented in Table 2. Experimental mass distribution data for $^{12}$C, $^{16}$O, and $^{30}$Si + $^{238}$U were collected from Refs. [34,10,5]. The mass ratio distributions below and above the respective fusion barrier energies are shown for comparison, and the relative yields are normalized. The GEMINI$^{++}$ predicted mass ratio distributions for all these reactions are shown as shaded areas in Fig. 5. As shown, the GEMINI$^{++}$ predicted mass distributions reproduce the experimental distribution reasonably well for light projectile systems ($^{11}$B, $^{12}$C, $^{16}$O, $^{19}$F), with the exception of the $^{30}$Si + $^{238}$U system, where the measured one significantly differ at below-barrier energy. For $^{30}$Si + $^{238}$U reaction, the measured mass distribution is predominantly asymmetric at lower energy, and single Gaussian fitting is not straightforward. In this reaction, the contribution of non-compound events is known to result in a larger mass variance or an asymmetric mass distribution, deviating from a single Gaussian shape [5]. The asymmetric component of the mass distribution in $^{30}$Si + $^{238}$U was attributed to the orientation-dependent quasi-fission process, which is dominant at below-barrier energies. Interestingly, the measured mass-ratio distribution at lower excitation energies for $^{249}$Bk and $^{257}$Md nuclei is well reproduced by GEMINI$^{++}$ predictions, while GEF exhibits slight deviations.

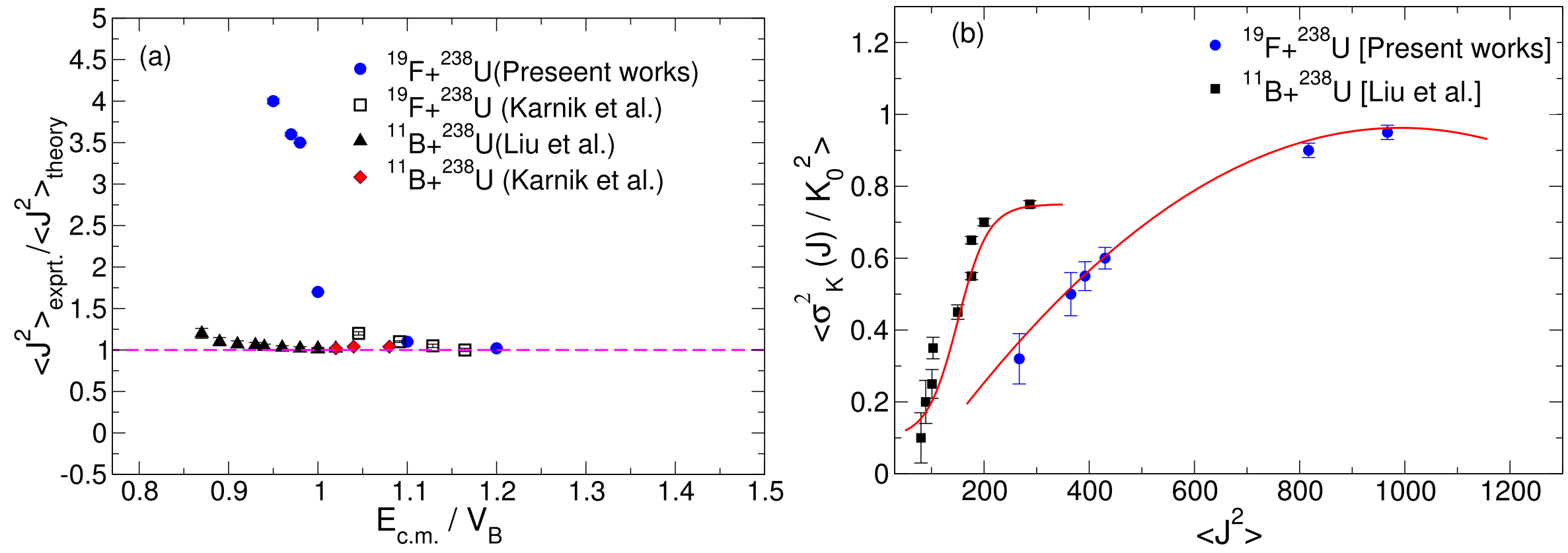

**Fig. 4.** (a) The average ratio $\langle J^2\rangle_{\text{expt}}/\langle J^2\rangle_{\text{theory}}$ as a function of $E_{\text{c.m.}}/V_B$. (b) The ratio $\frac{\sigma_K^2(J)}{K_0^2}$ plotted as a function of the mean-square angular momentum $\langle J^2\rangle$, calculated using Eqs. (1)-(4) for the actinide nuclei $^{249}$Bk and $^{257}$Md, produced in fusion reactions $^{11}$B [11] and $^{19}$F + $^{238}$U. The solid red curve corresponds to the theoretical prediction for the K relaxation model using Eq. (4).

**Table 2**
Entrance channel parameters for various reactions with projectiles on a $^{238}$U target. The table shows the projectile, compound nucleus (CN), asymmetry parameters ($\alpha$, $\alpha_{bg}$), fissility parameters ($\chi$, $\chi_{eff}$, $\chi_m$), $Z_P Z_T$, Bass barrier $V_B$, Q-value (negative, in MeV), $E_{cm}$ (MeV), and excitation energy $E^*$ (MeV).

| Reactions with different projectiles on $^{238}$U target | | | | | | | | | | | | |
|---|---|---|---|---|---|---|---|---|---|---|---|---|
| Projectile | CN | $\alpha$ | $\alpha_{bg}$ | $\chi_{CN}$ | $\chi_{eff}$ | $\chi_m$ | $Z_P Z_T$ | $V_B$ (MeV) | -Q (MeV) | $E_{cm}$ (MeV) | $E^*$ (MeV) | Ref. |
| $^{11}$B | $^{249}$Bk | 0.911 | 0.892 | 0.813 | 0.391 | 0.655 | 460 | 53.37 | 13.87 | 49.7 | 35.8 | Present work |
| | | | | | | | | | | 61.6 | 47.3 | |
| $^{12}$C | $^{250}$Cf | 0.904 | 0.896 | 0.823 | 0.393 | 0.680 | 552 | 64.22 | 23.86 | 60.9 | 36.9 | Yadav et al. [34] |
| | | | | | | | | | | 70.4 | 46.5 | |
| $^{16}$O | $^{254}$Fm | 0.874 | 0.902 | 0.841 | 0.462 | 0.715 | 736 | 84.37 | 38.33 | 82.3 | 38.5 | Hinde et al. [10] |
| | | | | | | | | | | 108 | 62.8 | |
| $^{19}$F | $^{257}$Md | 0.852 | 0.905 | 0.849 | 0.483 | 0.727 | 828 | 93.67 | 43.17 | 85 | 42 | Present work |
| | | | | | | | | | | 100 | 56.8 | |
| $^{30}$Si | $^{268}$Sg | 0.776 | 0.919 | 0.893 | 0.612 | 0.799 | 1288 | 139.7 | 93.92 | 129 | 35.5 | Nishio et al. [5] |
| | | | | | | | | | | 149 | 55.5 | |

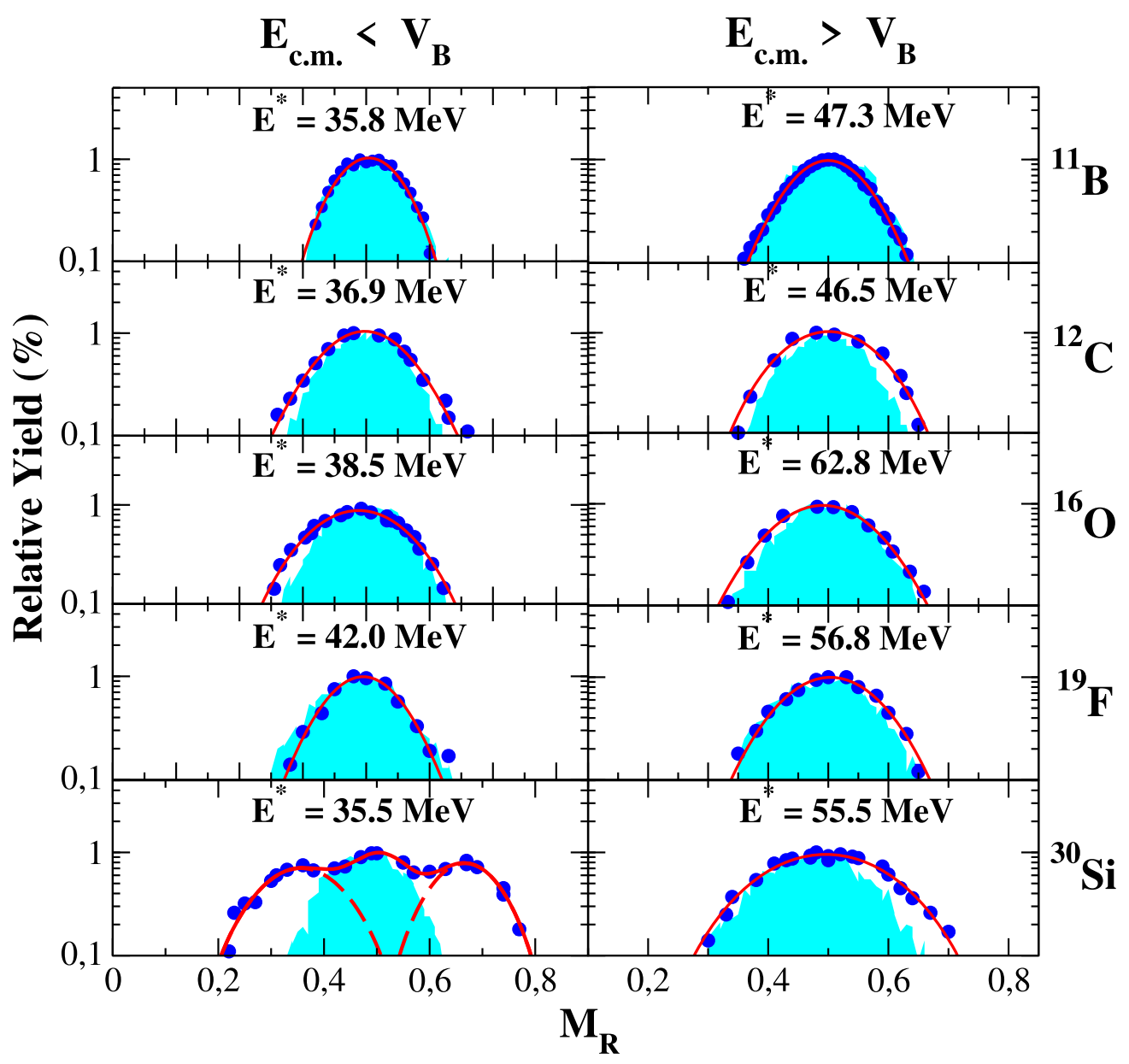

**Fig. 5.** Comparison of experimental mass-ratio spectra (blue circles) along with simulated (shaded) distributions for the reactions $^{11}$B, $^{12}$C, $^{16}$O, $^{19}$F, $^{30}$Si, + $^{238}$U. Right panel represents energies above respective barrier and left panel below barrier energies. The measured mass distribution data for $^{12}$C, $^{16}$O, $^{30}$Si + $^{238}$U reaction are taken from Yadav et al., Hinde et al., Nishio et al., Ref. [34,10,5]. The solid red curve corresponds to Gaussian representing the measured mass-ratio distributions. Shaded regions correspond to the theoretical prediction of GEMINI$^{++}$ [30]. The dashed red curve corresponds to the peaks of the lighter and heavier fragments in the fission of $^{268}$Sg nuclei. The mass distribution data have been normalized.

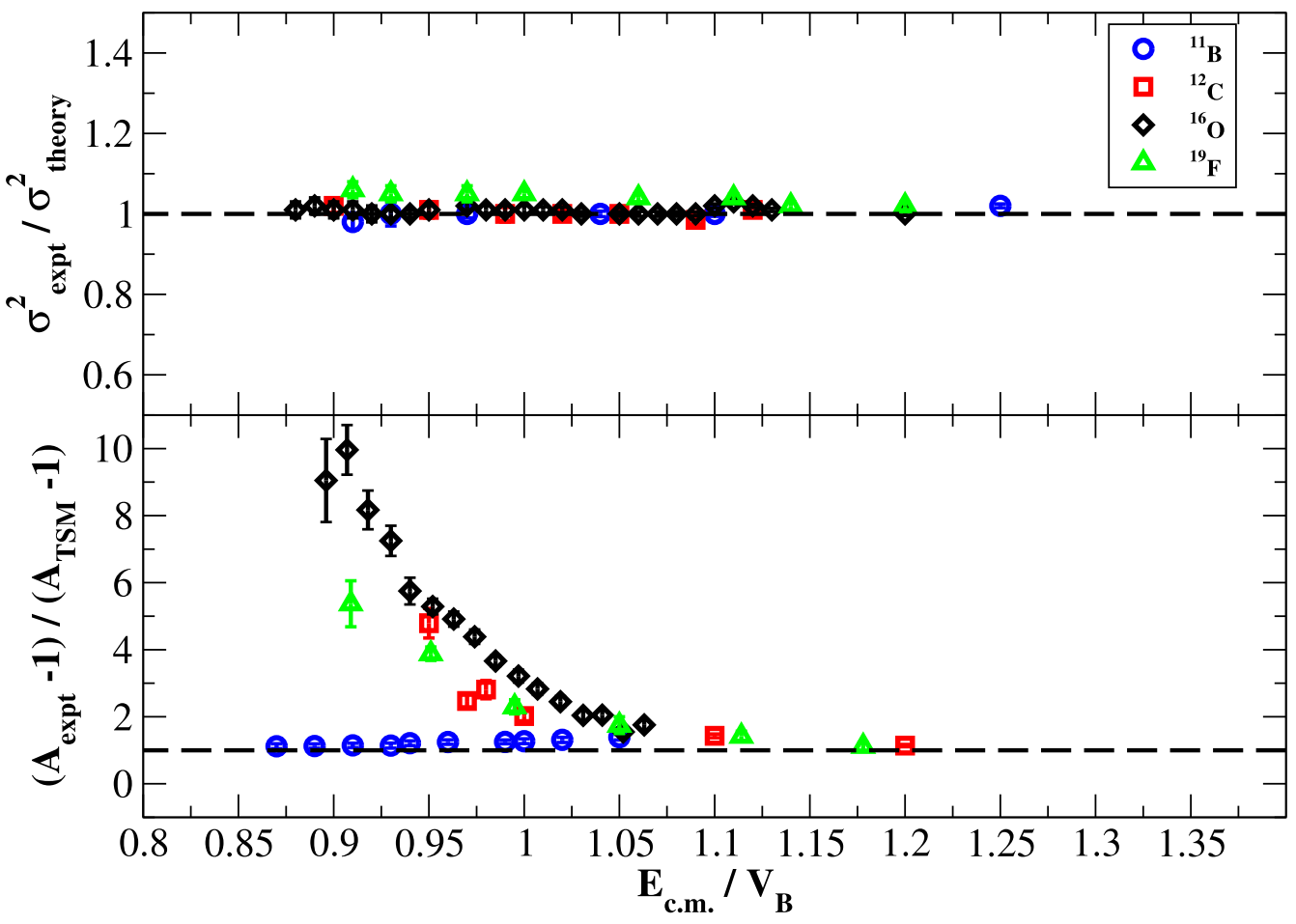

**Fig. 6.** Upper panel: Ratio of experimental mass-ratio variance ($\sigma^2_{\text{expt.}}$) against the GEMINI++ predictions ($\sigma^2_{\text{theory}}$) as a function of $E_{\text{C.M.}}/V_B$. Lower panel: Ratio of experimental values of FF angular anisotropies ($A_{\text{expt}} - 1$) with respect to TSM model ($A_{\text{TSM}} - 1$) predictions as a function of $E_{\text{c.m.}}/V_B$, for the four reactions $^{11}$B, $^{12}$C, $^{16}$O, $^{19}$F + $^{238}$U. The measured mass distribution data for $^{12}$C, $^{16}$O, + $^{238}$U are taken from Yadav et al. and Hinde et al. (Refs. [34,10]), and the angular anisotropy data for $^{11}$B, $^{12}$C, and $^{16}$O are taken from Refs. [10–12].

The ratio of experimental mass variance extracted from single Gaussian fit and the mass variance given by GEMINI++ predictions were calculated, and the trend of the estimated ratios is shown in Fig. 6 (Upper panel). As observed, no anomalous increase in mass variance below fusion barrier energies is detected for these reactions; the experimental data and GEMINI$^{++}$ predictions show close agreement. This suggests that GEMINI$^{++}$ adequately describes the observed fission fragment mass distributions resulting from the fission of heavy $^{257}$Md actinide nuclei or neighboring nuclei produced by $^{11}$B, $^{12}$C, $^{16}$O, and $^{19}$F + $^{238}$U reactions. Hinde et al. conducted a quantitative analysis of the asymmetric components due to "fast quasi-fission" in fission induced by projectiles $^{24}$Mg, $^{28,30}$Si, and $^{34,36}$S on $^{232}$Th and $^{238}$U targets [14]. For these systems, in addition to asymmetric components, fission centered on mass symmetry exhibits wider mass widths and large angular anisotropies, indicating the presence of quasi-fission. In Fig. 6 (Lower panel), we show the trend of measured anisotropies as a function of $E_{C.M.}/V_B$ for reactions induced on $^{238}$U, similar to the systematic given by Hinde et al. [14]. The experimental data for $^{11}$B, $^{12}$C, and $^{16}$O have been taken from Hinde et al., Liu et al., Lestone et al. [10–12]. It is shown that except for reaction $^{11}$B + $^{238}$U in which the anisotropies closely follow the TSM values at all measured energies [11], anisotropies measured for other reactions follow similar trend; anisotropy value increases below the fusion barrier in contrast to TSM predictions. Hinde et al. reported systematic behavior of anisotropies based on the assumption that QF anisotropy may be similar for all reactions and thus estimated the fraction of slow QF events in each reaction [14].

Thus the studies of mass and angular distributions suggest that if the $K$ relaxation time is short compared to the fission timescale, the nucleus is more likely to lose memory of its initial $K$ value, leading to symmetric or slightly asymmetric fission fragment distributions. Conversely, if the $K$ relaxation time is long, the nucleus may retain its initial configuration, resulting in more asymmetric fission modes. This points to an interplay between $K$ relaxation and deformation-dependent dissipation. To better understand fission dynamics in heavy actinide nuclei, the relationship between $K$ relaxation, dynamic dissipation, and shell corrections should be further explored. The Langevin dynamical model offers a promising approach for investigating this mass region. For example, recent studies using pre-scission neutron multiplicity to estimate fission timescales suggest that dissipation strongly influences fusion-fission dynamics in these nuclei [40].

## 4. Summary

The experimental fission fragments mass-ratio distributions in the fission of $^{249}$Bk and $^{257}$Md heavy actinide nuclei, produced via $^{11}$B and $^{19}$F + $^{238}$U reactions, along with the angular distributions of $^{257}$Md nuclei, have been measured and analyzed within statistical model GEF, GEMINI++, TSM framework. The FF mass distributions for these nuclei exhibit the presence of an asymmetric fission mode around 50 MeV and beyond, excitation energies, suggesting multichance fission. However, it is difficult to disentangle the contributions of the asymmetric and symmetric superlong, supershort fission modes based on the experimental data alone. Further investigation into the non-compound nucleus process, comparing experimental mass-ratio distributions with the Monte Carlo GEMINI$^{++}$ model for these nuclei, shows good agreement with predictions.

From comparative studies of reactions induced by projectiles $^{11}$B, $^{12}$C, $^{16}$O, $^{19}$F, and $^{30}$Si on the deformed $^{238}$U target, it appears that fission fragments mass-ratio distribution measurements exhibit nearly all signatures of complete fusion-fission events around the capture barrier energies in reactions induced by projectiles with $A_p < 22$. For reactions induced by projectiles with $A_p > 22$ on similar actinide targets, signatures of non-compound events are clearly visible, manifested as asymmetric and broad mass distributions at sub-barrier energies. The observed discrepancy between measured angular anisotropy in $^{19}$F + $^{238}$U and the predictions of the transition state model supports the non-equilibrium aspects of fusion-fission dynamics. The role of deformation-aligned quasi-fission at sub-barrier energies is more evident in reactions involving heavier projectiles on deformed targets, where large anisotropies and mass variance are observed.

Many hypotheses and theoretical understandings are still being discussed regarding the timescale of the fission process, particularly concerning slow QF in these nuclei. However, the most conclusive way to confirm QF is by measuring the evaporation residue cross-section, pre-scission neutron multiplicities; without this measurement, it remains difficult to definitively identify any non-equilibrium fission processes in the fission of $^{249}$Bk and $^{257}$Md actinide nuclei produced by $^{11}$B and $^{19}$F projectiles. Even though the idea of multi-mode fission occurring due to shell closure around the mass region 140 for heavy fragments and 96 for lighter fragments for fission of in these nuclei has been proposed, it is still insufficient to fully explain the fission fragment mass distribution observations. Thus, a complex interplay between $K$ relaxation time, dynamic dissipation, and its influence on shell correction, and consequently the outcome of the fission process in terms of fission modes in heavy actinide nuclei, might need to be explored for a better understanding of the fission dynamics of these nuclei.

## Declaration of competing interest

The authors declare that they have no known competing financial interests or personal relationships that could have appeared to influence the work reported in this paper.

## Acknowledgements

We thank the IUAC accelerator group members for providing us stable pulsed beams throughout the experiment. Author acknowledges Prof H. J. Wollersheim of GSI for providing us the target.

## Data availability

Data will be made available on request.